\UseRawInputEncoding

\documentclass[conference]{IEEEtran}

\usepackage[dvips]{graphicx}

\ifCLASSINFOpdf
\else
\fi
\begin{document}
%
\title{A Study to Optimize Heterogeneous Resources for Open IoT}

\author{\IEEEauthorblockN{Yoji Yamato, Naoto Hoshikawa, Hirofumi Noguchi, Tatsuya Demizu and Misao Kataoka}
\IEEEauthorblockA{NTT Network Service Systems Laboratories\\
NTT Corporation\\
Musashino-shi, Tokyo 180--8585, Japan\\
Email: yamato.yoji@lab.ntt.co.jp}}


%


\maketitle

\begin{abstract}
Recently, IoT technologies have been progressed, and many sensors and actuators are connected to networks. Previously, IoT services were developed by vertical integration style. But now Open IoT concept has attracted attentions which achieves various IoT services by integrating horizontal separated devices and services. For Open IoT era, we have proposed the Tacit Computing technology to discover the devices with necessary data for users on demand and use them dynamically. We also implemented elemental technologies of Tacit Computing. In this paper, we propose three layers optimizations to reduce operation cost and improve performance of Tacit computing service, in order to make as a continuous service of discovered devices by Tacit Computing. In optimization process, appropriate function allocation or offloading specific functions are calculated on device, network and cloud layer before full-scale operation.
\end{abstract}



%
\IEEEpeerreviewmaketitle

\section{Introduction}
Recently, IoT (Internet of Things) technologies have been progressed (e.g., \cite{Sound}-\cite{ICGIP}), and many IoT devices are connected to networks. IoT application areas are wide such as manufacturing, supply chain, maintenance which Industrie4.0 targets and also health care, agriculture and energy. 

However, current IoT applications tend to be one off solutions with vertical integration for specified targets. System integration businesses which carry out sensor selecting, analyzing, visualizing and taking actions together are popular for IoT applications. Therefore, each IoT application needs high development cost. 

To reduce costs of various IoT application development and operation, shared use of devices such as sensors and actuators and integrating horizontal separated devices and services are needed. This approach is called Open IoT and attracts attentions for future IoT applications acceleration.

We have proposed Tacit Computing technology \cite{Tacit} to utilize shared use of devices from various services for Open IoT era. Tacit Computing can discover devices with necessary data for users on demand based on live data of each device, and can use them. We have also implemented elemental technologies of Tacit Computing.

However, we think discovering and using the device based on the situation at that time only answers the user's needs only one time. In order to make it a service that keeps answering to the needs of user, it is needed to provide the continuous use of devices with necessary data for the user at a reasonable price. Therefore, in this paper, we propose three layers optimizations to reduce operation cost and improve performance so as to make a continuous service with devices which Tacit Computing discovers and uses. In optimization processing, appropriate function allocations and offloading specific processing are conducted on device, network and cloud layers before full-scale operation. This paper explains approaches of three layers optimizations and we report detail evaluation of each optimization technology in other papers.

\section{Outline of Tacit Computing}
Tacit Computing answers users' requests by discovering and coordinating appropriate resources for users from cloud layer, network layer and device layer (Figure 1). Tacit Computing has three layers structure but its concept is to process as much as possible in the device layer which is close to user site to cope with the situation changing. By processing within lower layers, Tacit Computing can reduce network traffic amount and prevents leak of high privacy data. Here, we review live data discovering technology and device virtualization technology which are key technologies for Open IoT era.

Live data discovering technology discovers devices which provide necessary data for users. For example, there is an existing IoT application in which multiple sensors are attached to facilities such as bridges and degraded states are monitored continuously. In this case, it can not be thought that the facility degradation is going soon. Therefore, multiple points sensor data are collected to the cloud with a period of several hours, and the cloud analyzes the change of degraded state by statistical software or machine learning techniques. On the other hand, in an example where information guidance or warning alert are performed on a person appearing on a fixed point camera, only a few seconds the person appears in the camera, and only the image of the camera is meaningful to the person. In this way, data occurs in device layer and changes from moment to moment is called live data. To discover live data which is necessary for users, live data discovering technology distributes analyzing functions to lower layers instead of waiting for data to come up to cloud layer.

For example, we suppose a situation that a user's friend participates a marathon contest and a user wishes to see movies of cameras that the friend appears. In this case, bib number of the friend is requested for searching key to Tacit Computing. Tacit Computing distributes image analyzing functions such as OpenCV to gateways or network edges which accommodates cameras, and the functions analyze movies, extracts bib number of the friend by image analysis, then Tacit Computing can discover the camera which the friend appears at that time.

Next, Tacit Computing needs to use the device after it discovers the device with necessary user data. Because IoT devices are developed by many companies and each device has different protocol, interface, address and so on. Therefore, device virtualization technology conceals differences of each device interface and so on. In the above example, the method of use varies for each camera. Tacit Computing converts each request from common requests such as camera image acquisition to specific requests for individual camera by each device adapter on the gateways or network edges. By this method, users can use devices without considering each device difference. When we use devices, we can also use Semantic Web Services  or other methods (e.g., \cite{ICWS}-\cite{INTECH}).

 \begin{figure}[tb]
 \begin{center}
  \includegraphics[width=90mm]{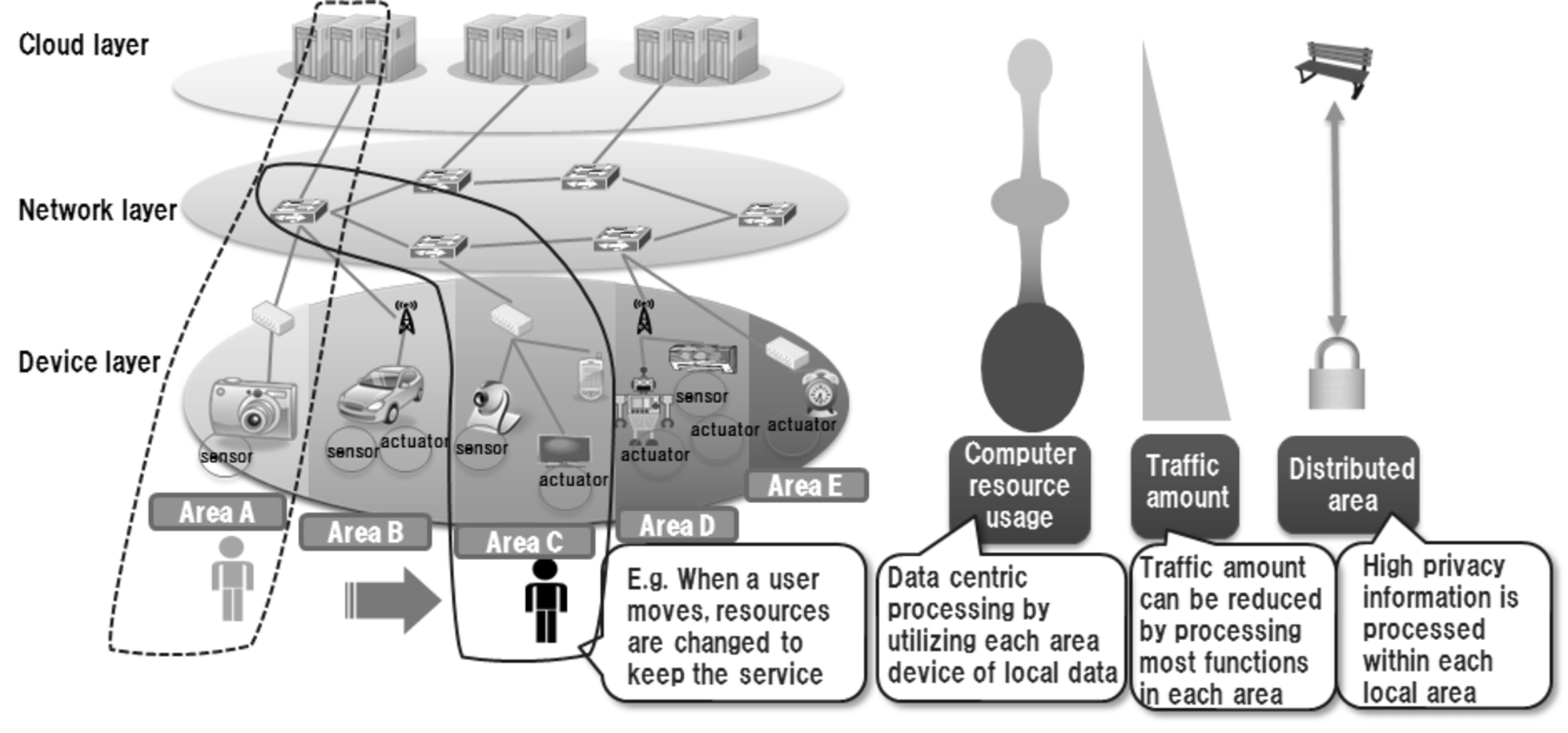}
 \end{center}
 \caption{Outline of Tacit Computing}
 \end{figure}

\section{Problems to be solved}
Tacit Computing enables ad-hoc devices using for users by live data discovering technology and device virtualization technology. However, we think discovering and using the device based on the situation at that time by previous Tacit Computing only answers the user's needs only one time. 

For example of Tacit Computing, let us consider tracking cameras. Tracking cameras are usage that movies of small children in schools or roads are taken by security cameras near the children and parents can see the movies by their mobile terminals. Tracking cameras can satisfy parents' needs to confirm children's safety at that time, but we think it is difficult to accept if it is charged as a certain price as once movie checking. We think it can be acceptable that it is charged as a monthly fee of continuous monitoring service in which parents can see movies of their children when they want to see on mobile terminals, but children are watched by machine learning or so on even when parents do not see and alerts are sent to parents in case of anomaly such as when a suspicious person approaches to children on roads.

As a processing of tracking cameras, Tacit Computing discovers the camera in which the child appears, and delivers movies of the camera to the parents' mobile terminals when the parents request movies. And for watching, image analyzing functions such as OpenCV library are arranged on gateways or network edge SSE (Subscriber Service Edge) which accommodate cameras and image analyzing functions analyze images from the camera movie in which the child appears. The result of image analysis is summarized to the feature vector and it is aggregated to the cloud, and using cloud technologies (e.g., \cite{IEEE}-\cite{Heat}), anomaly such as suspicious person approaching is analyzed by machine learning techniques (e.g., Local Outlier Factor). When anomaly is detected, alerts are sent to parents.

Tracking cameras can be achieved by not only Tacit Computing but also solution services using security cameras and machine learning techniques. Therefore, we set a problem to be solved as "providing continuous services that use devices discovered by Tacit Computing at reasonable prices".

In Tacit Computing, when we discover and use a device ad-hoc, we need to be able to use devices first and costs and performances are ignored. However, in order to provide it reasonably as a continuous service, it is necessary to reduce operation cost and improve performance.

Therefore, in this paper, we propose optimizations of appropriate function allocation and offloading specific processing in device, network and cloud layers. Optimizations are supposed to be performed during the first ad-hoc device usage period. 

\section{Proposal of three layers optimizations}
In this section, we propose optimization approaches to reduce operation cost and improve performance for continuous service with device using of Tacit Computing. 

In device layer, it is necessary to switch a device that satisfies the needs of user at first. In the case of tracking cameras, it means to select the camera in which the image of the child appears based on location of the child. At this time, if we analyze images with all cameras connected to the network and use only the camera with the child appears, the cost will be very high. Thus, it is necessary to narrow down cameras near the child by identifying children with some identifiers and using cameras' metadata of location. Since many IoT devices are non-IP devices, we are considering to analyze communication patterns of IoT devices including non-IP devices and to assign metadata automatically or semi-automatically from the communication pattern.

Furthermore, in device layer, where to allocate analyze function on a gateway from gateways which accommodate devices affects operation costs much. In the case of tracking cameras, an image analysis function such as OpenCV is used. We are considering that image analysis functions are arranged in the gateways of the areas where children often pass, and when children move beyond that areas, image analysis functions are distributed at that times.

In network layer, as well as function placement for the gateways, we are considering optimizing function placement on network edges such as SSE. Because the caching and arrangement of information on the network is studied by information centric network, we consider to apply those ideas. 

Depending on the service characteristic, bandwidth reservations, priority controls or other setting may be necessary in the network. For example, in the case of NGN (Next Generation Network), there is a function for reserving bandwidth called RACF (Resource and Admission Control Function), and such network side setting preparation also needs to be performed during the optimization processing period.

In cloud layer, where to process in the cloud greatly affects cost and performance. Firstly, we deploy the processing function to the cloud of the DC (Data Center) that has a small delay from the network edge that accommodates devices frequently used. Furthermore, since the size of the cloud resource also affects the operation cost, its resource size is also subject to optimization. For example, in the case of tracking cameras, the cloud analyzes summarized feature vectors by machine learning techniques and alerts suspicious person and so on. To detect anomaly in real time, stream processing functions such as Storm or Spark are used, so the cloud keeps appropriate resource sizes for these functions.

In addition, in recent days, servers with heterogeneous hardware such as GPU and FPGA are increasing in the clouds. By utilizing them, it is possible to achieve high performance by offloading matrix calculation and so on to GPU and offloading specific processing such as FFT (Fast Fourier Transformation) calculation to FPGA.

Furthermore, although we have described each of device layer, network layer and cloud layer so far, depending on the usage style of services, appropriate function placement is changed. Therefore, by modeling the cost (server cost for reserving resources and network cost based on generated traffic) and performance (throughput and delay) when each processing is processed in each layer, we are considering to calculate appropriate function placement. It is a combination optimization problem in order to optimize the performance with the price/cost for users as the constraint conditions, thus we adopt a method with a certain time calculation to find a certain degree of solution during limited optimization period.

\end{document}